\documentclass{article}
\usepackage{hyperref}
\usepackage{amsmath}
\usepackage[margin=1in, footskip=0.25in]{geometry}
\usepackage{xcolor}
\usepackage{graphicx}
\graphicspath{ {./figures/} }
\usepackage{float}
\usepackage{setspace}
\usepackage{subcaption}
\usepackage[style=ieee]{biblatex}
\begin{document}

\Large
 \begin{center}
Bayesian Prevalence Estimation from Pooled and Individual Data\\ 

\hspace{10pt}

\large {Matthew Ritch and Charles Copley} \\
\hspace{10pt}
\small  \textit{Ancera Data Science}

\end{center}

\normalsize

\noindent\sc{Abstract} \normalfont Pooled and individual disease testing are common methods for determining the population prevalences of diseases. Recently, researchers have used Monte Carlo Markov Chain methods to estimate population prevalence from the combined streams of these two types of testing data. We propose an analytical solution for estimating population prevalence from combined individual and pooled binary sampling data. We also use simulated sampling data to characterize these posterior distributions under a variety of sampling conditions, including a range of true prevalences, variable numbers of pooled and individual tests, variable number of individual samples per pooled sample, and a range of values for test sensitivity and specificity. 
\section{Introduction}
Binary tests are frequently used to measure disease prevalence in a large population. When test kits are scarce or expensive, multiple individual samples can be combined for a pooled or group test to save resources \cite{du_combinatorial_2000}. Recently, researchers have investigated using Bayesian data fusion \cite{allen_statistical_2017} methods for using pooled and individual test data together to estimate population prevalences via Monte Carlo Markov Chain (MCMC) \cite{hoegh_estimating_2021}\cite{scherting_pool_2021}\cite{scherting_estimation_2023}.

We have developed an analytical Bayesian method for combining data from both individual and pooled tests into a single analytical posterior distribution for the true population prevalence, obviating the need for MCMC methods and thereby saving compute time and resources. As we will show, pooled testing is more useful at low true prevalences. We present simulation results for a variety of sampling conditions. These results can be used to inform sampling program design given a test budget and a preliminary estimate of disease prevalence.

\section{The Posterior Probability Distribution for Population Prevalence}

\subsection{Definitions}
\begin{noindent}
\\Define $P$ as a random variable for population prevalence. We will use $p$ to denote outcome values of that random variable.\\
Define $m$ as the number of individual tests conducted.\\
Define $Y_1,Y_2,...Y_m$ as random variables for the binary results of these individual tests.\\
Define $y$ as the number of observed positive individual tests.\\
Define $n$ as the number of pooled tests conducted.\\
Define $Z_1,Z_2,...Z_n$ as random variables for the binary results of these pooled tests.\\
Define $z$ as the number of observed positive pooled tests.\\
Define $\pi_q$ as the probability of a pooled sample testing positive when $q$ individuals are pooled.\\
\end{noindent}

We assume that each individual sample's disease status is identically and independently distributed and is positive with probability $P$. This allows us to derive simple expressions for the probability of a positive pooled test and the joint probability of our individual and pooled testing results. This assumption is also made in Hoegh et al. 2021 \cite{hoegh_estimating_2021}, where they explain that “Implicitly the calculation of $\pi$ in Equation 1 assumes that the samples are independent. For the applications we are primarily focused on, viral surveillance in wildlife populations where individual samples can be randomly assigned to pools, this is usually a reasonable assumption.”

The probability of a pooled sample testing positive when $q$ individuals are pooled, $\pi _q$, is equivalent to the probability that at least one of the samples which are combined into the pooled sample is positive.

\[ \pi _q = 1 - Pr(all\;of\;the\;samples\;are\;negative) \]

so

\[ \pi _q  = 1 - (1 - p)^q \]

Each $Y_i$ and $Z_i$ are i.i.d. Bernoulli
\[ Y_i \sim Bernoulli(p)\]
\[ Z_i \sim Bernoulli(\pi _q)\]
Define random variables $Y$ and $Z$ as 
\[ Y = \sum_{i=1}^{m}{Y_i}\]
\[ Z = \sum_{i=1}^{n}{Z_i}\]
Then $Y$ and $Z$ follow binomial distributions and are independent.
\[ Y \sim Binomial(m,p)\]
\[ Z \sim Binomial(n,\pi _q)\]
so
\begin{equation}\label{indivBin} Pr(Y=y|P=p) = \binom{m}{y}    (p)^{y} (1-p)^{m-y} \end{equation}
\begin{equation}\label{poolBin}Pr(Z=z|P=p) = \binom{n}{z}     (\pi _q)^{z} (1-\pi _q)^{n-z} \end{equation}
In general, we assume a beta prior distribution for $P$
\[ P \sim Beta(\alpha, \beta)\]
so
\begin{equation}\label{prior} Pr(P=p) = \frac{ p^{\alpha - 1} (1-p)^{\beta - 1}}{B(\alpha, \beta)}\end{equation}
where $B(\alpha, \beta)$ is the beta function of $\alpha$ and $\beta$.\\
For the simulation study, we use the uniform prior
\[ \alpha = 1, \beta = 1\]
Bayes' Theorem can be stated for this problem as
\begin{equation}\label{bayes} Pr(P=p|Y=y, Z=z) = \frac{Pr(Y=y, Z=z|P=p)Pr(P=p)}{Pr(Y=y, Z=z)} \end{equation}
We will solve for $Pr(P=p|Y=y, Z=z)$.

\subsection{$Pr(Y=y, Z=z|P=p)Pr(P=p)$}
Because we have assumed that each sample is i.i.d. $Bernoulli(P)$, 
\[Pr(Y=y, Z=z|P=p) = Pr(Y=y|P=p) Pr( Z=z|P=p) \]
Substitute in equations \ref{indivBin} and \ref{poolBin}:
\[= \binom{m}{y}  \binom{n}{z}  (p)^{y} (1-p)^{m-y}  (\pi _q)^{z} (1-\pi _q)^{n-z}\]
and recall our assumed prior for $P$, equation \ref{prior}
\[ Pr(P=p) = \frac{ p^{\alpha - 1} (1-p)^{\beta - 1}}{B(\alpha, \beta)}\]
so
\begin{equation}\label{step0} Pr(Y=y, Z=z|P=p)Pr(P=p) =  \frac {\binom{m}{y}  \binom{n}{z}  (p)^{y} (1-p)^{m-y}  (\pi _q)^{z} (1-\pi _q)^{n-z} p^{\alpha - 1} (1-p)^{\beta - 1}} {B(\alpha,\beta)}  \end{equation}
combine terms
\begin{equation}\label{step1} Pr(Y=y, Z=z|P=p)Pr(P=p)= \frac {\binom{m}{y}  \binom{n}{z}  (p)^{y+\alpha - 1} (1-p)^{m-y + \beta - 1}  (\pi _q)^{z} (1-\pi _q)^{n-z}} {B(\alpha,\beta)}  \end{equation}
recall
\[ \pi _q  = 1 - (1 - p)^q \]
so 
\begin{equation}\label{1minus}  (1 - \pi _q)^{n-z} = ((1-p)^{q})^{n-z} = (1-p)^{qn-z q}\end{equation}
By binomial expansion:
\begin{equation}\label{pibin}  \pi _q ^{z} = [1 - (1-p)^{q}]^{z} = \sum _{i=0} ^{z} \binom{z}{i} (-1)^{i} (1-p)^{qi}\end{equation}
Substitute these expressions for $(1 - \pi _q)^{n-z}$ and $\pi _q ^{z}$ in equations \ref{1minus} and \ref{pibin} into equation \ref{step1}
\[Pr(Y=y, Z=z|P=p)Pr(P=p) = \frac {\binom{m}{y}  \binom{n}{z}  (p)^{y+\alpha - 1} (1-p)^{m-y + \beta - 1 +qn -q z} [\sum _{i=0} ^{z} \binom{z}{i} (-1)^{i} (1-p)^{qi}]} {B(\alpha,\beta)}  \]
Move terms inside the summation:
\[ Pr(Y=y ,Z=z |P=p)Pr(P=p)) = \frac{\binom{m}{y} \binom{n}{z}}{B(\alpha,\beta)} \sum ^{z} _{i=o} \binom{z}{i} (-1)^{i} p^{y + \alpha - 1} (1 - p)^{m - y + \beta - 1 + qn -q z} (1-p)^{qi} \]
Combine terms to find that
\begin{equation}\label{joint} Pr(Y=y ,Z=z |P=p)Pr(P=p)) = \frac{\binom{m}{y} \binom{n}{z}}{B(\alpha,\beta)} \sum ^{z} _{i=o} \binom{z}{i} (-1)^{i} p^{y + \alpha - 1} (1 - p)^{m - y + \beta - 1 + qn -q z +qi} \end{equation}

\subsection{$Pr(Y=y , Z=z)$}
Now the joint probability $Pr(Y=y , Z=z, P=p)$ can be expressed as
\[Pr(Y=y , Z=z |P=p)Pr(P=p)\]
So $Pr(Y=y , Z=z)$ can be expressed as 
\[ Pr(Y=y , Z=z) = \int _0 ^1 Pr(Y=y , Z=z, P=p) dp \]
\[ = \int _0 ^1 Pr(Y=y , Z=z | P=p)Pr(P=p) dp \]
so, using equation \ref{joint},
\[ Pr(Y=y , Z=z) =  \frac{\binom{m}{y} \binom{n}{z}}{B(\alpha,\beta)} \int _0 ^1  \sum ^{z} _{i=o} \binom{z}{i} (-1)^{i} p^{y + \alpha - 1} (1 - p)^{m - y + \beta - 1 + qn -q z +qi}dp\]
By linearity of integration:
\[ Pr(Y=y , Z=z) =  \frac{\binom{m}{y} \binom{n}{z}}{B(\alpha,\beta)} \sum ^{z} _{i=o} \binom{z}{i} (-1)^{i}  \int_{0}^{1}  p^{y + \alpha - 1} (1 - p)^{m - y + \beta - 1 + qn -q z +qi}dp\]
By definition, 
\[ \int_{0}^{1}p^{a-1}(1-p)^{b-1} = B(a,b) \] 
where $B(a,b)$ is the beta function of $a$ and $b$, so
\begin{equation}\label{margined} Pr(Y=y , Z=z) =  \frac{\binom{m}{y} \binom{n}{z}}{B(\alpha,\beta)} \sum ^{z} _{i=o} \binom{z}{i} (-1)^{i} B(y +\alpha, m - y + \beta + qn -q z +qi) \end{equation}

\subsection {The Posterior Distribution for $P$}
For simplicity of notation, make these two substitutions
\[ \gamma = y + \alpha \]
\[ \delta = m - y + \beta +qn -qz \]
in equation \ref{margined} 
\[ Pr(Y=y , Z=z) =  \frac{\binom{m}{y} \binom{n}{z}}{B(\alpha,\beta)} \sum ^{z} _{i=o} \binom{z}{i} (-1)^{i} B(\gamma, \delta + qi)\]
and in equation \ref{joint}
\[Pr(Y=y ,Z=z |P=p)Pr(P=p)) = \frac{\binom{m}{y} \binom{n}{z}}{B(\alpha,\beta)} \sum ^{z} _{i=o} \binom{z}{i} (-1)^{i} p^{\gamma-1} (1 - p)^{\delta + qi - 1} \]
Recall Bayes' Theorem
\begin{equation}\tag{\ref{bayes} revisited} Pr(P=p|Y=y, Z=z) = \frac{Pr(Y=y, Z=z|P=p)Pr(P=p)}{Pr(Y=y, Z=z)} \end{equation}
so 
\begin{equation}\label{soln1}Pr(P=p|Y=y, Z=z)  = \frac {\sum_{i=0}^{z}  \binom{z}{i} (-1)^{i} p^{\gamma - 1} (1 - p)^{\delta + qi - 1} } {\sum_{i=0}^{z}  \binom{z}{i} (-1)^{i} B(\gamma, \delta + qi)}   \end{equation}
This is an analytical form of the posterior probability distribution for the population prevalence $P$.
Now let $f(p, \gamma, \delta + qi)$ be the PDF of the beta distribution with parameters  $\gamma$, and $\delta + qi$, evaluated at $p$.
\[ f(p, \gamma, \delta + qi) = \frac{p^{\gamma-1}(1-p)^{\delta + qi-1}} { B(\gamma, \delta + qi) }\]
So we can also write the posterior probability distribution for $P$ in this form:
\begin{equation}\label{soln2} Pr(P=p|Y=y, Z=z)  =  \frac {\sum_{i=0}^{z}  \binom{z}{i} (-1)^{i} B(\gamma, \delta + qi) f(p, \gamma, \delta + qi)} {\sum_{i=0}^{z}  \binom{z}{i} (-1)^{i} B(\gamma, \delta + qi)}  \end{equation}
This form is less useful computationally, but it demonstrates that the posterior distribution for $P$ can be viewed as a weighted sum of $1+z$ beta distributions. 

\subsection{Posterior Distribution for $P$ with Point Estimates for Sensitivity and Specificity}\label{sesp}
Now we will consider what happens to the posterior distribution when our binary testing is sometimes incorrect. We define sensitivity $s_e$ and specificity $s_p$ as 
\[ s_e = Pr(test\;is\;positive\;|\;individual\;is\;positive)\]
\[ s_p = Pr(test\;is\;negative\;|\;individual\;is\;negative)\]
For simplicity, we will assume without loss of generality that these values are the same for pooled tests. Thus, 
\[ Pr(pooled\;test\;is\;positive\;|\;at\;least\;one\;pooled\;individual\;is\;positive) = s_e\]
\[ Pr(pooled\;test\;is\;negative\;|\;all\;pooled\;individuals\;are\;negative) = s_p\]
We will also assume that we know the true values of $s_e$ and $s_p$. We can now modify our previous expression for $Pr(Y=y, Z=z|P=p)Pr(P=p)$. Again,
\[Pr(Y=y, Z=z|P=p) = Pr(Y=y|P=p) Pr( Z=z|P=p) \]
Now, we will modify \ref{indivBin} and \ref{poolBin} to account for sensitivity and specificity.
\begin{equation}\label{indivBinImpStart}
Pr(Y=y|P=p) = \binom{m}{y}    [s_e p + (1 - s_p)(1-p)]^{y} [(1-s_e)p+s_p (1-p)]^{m-y}
\end{equation}
\begin{equation}\label{poolBinImpStart}
Pr(Z=z|P=p) = \binom{n}{z}    [s_e \pi _q + (1 - s_p)(1-\pi _q)]^{z} [(1-s_e)\pi _q + s_p (1-\pi _q)]^{n-z} 
\end{equation}
As before, we will use the binomial theorem to make these expressions more amenable to the integration needed to find an analytical posterior distribution for $P$. First, look at \ref{indivBinImpStart}.
\[ [s_e p + (1 - s_p)(1-p)]^{y} = \sum_{i=0}^{y}\binom{y}{i} [s_e p]^{i} [(1 - s_p)(1-p)]^{y - i}  = \sum_{i=0}^{y}\binom{y}{i} s_e ^{\;i} (1 - s_p)^{y - i} p^{i} (1-p)^{y - i}\]
and
\[ [(1-s_e)p+s_p (1-p)]^{m-y} = \sum_{j=0}^{m-y} \binom{m-y}{j}[(1 - s_e)p]^{j} [s_p (1-p)]^{m-y-j} = \sum_{j=0}^{m-y} \binom{m-y}{j} (1 - s_e)^j s_p ^{\;m - y - j} p^{j} (1-p)^{m - y - j} \]
so
\begin{equation}\label{indivBinImp}
Pr(Y=y|P=p) =  \binom{m}{y} \sum_{i=0}^{y}\sum_{j=0}^{m-y}\binom{y}{i} \binom{m-y}{j}
s_e^{\;i}(1-s_e)^{j} s_p^{\; m -y-j} (1-s_p)^{y - i} 
p^{i+j}(1-p)^{m - i - j}
\end{equation}
Similarly, \ref{poolBinImpStart} can be modified to
\begin{equation}\label{poolBinImp}
Pr(Z=z|P=p) = \binom{n}{z} \sum_{k=0}^{z}\sum_{l=0}^{n-z} \binom{z}{k} \binom{n-z}{l}
s_e^{\;k}(1-s_e)^{l} s_p^{\; n -z-l} (1-s_p)^{z - k} 
\pi_q^{\;k+l}(1-\pi_q)^{n - k - l}
\end{equation}
Define $g(i,j,k,l)$ as
\begin{equation}\label{g}
g(i,j,k,l) = \binom{y}{i} \binom{m-y}{j} \binom{z}{k} \binom{n-z}{l}
s_e^{\;i+k}(1-s_e)^{j + l} s_p^{\; m -y-j +  n -z-l} (1-s_p)^{y - i + z - k} 
\end{equation}
Combining \ref{indivBinImp} and \ref{poolBinImp} with our prior \ref{prior},
\[ Pr(Y=y|P=p)Pr(Z=z|P=p)Pr(P=p) = \]
\begin{equation}\label{impJointPi}
\begin{split}
\frac{\binom{m}{y}\binom{n}{z}}{B(\alpha,\beta)} \sum_{i=0}^{y}\sum_{j=0}^{m-y} \sum_{k=0}^{z}\sum_{l=0}^{n-z} 
g(i,j,k,l) p^{i+j + \alpha - 1}(1-p)^{m - i - j + \beta - 1} \pi_q^{\;k+l}(1-\pi_q)^{n - k - l}
\end{split}
\end{equation}
To continue, we can express $(1 - \pi_q)^{m - k - l}$ and $\pi_q ^{k+l}$ in terms of $p$, as we did in equations \ref{1minus} and \ref{pibin}.
recall
\[ \pi_q = 1 - (1-p)^q\]
so 
\[ (1 - \pi_q)^{n - k - l} = ((1-p)^{q})^{n-k-l}=(1-p)^{nq-kq-lq} \]
and
\[ \pi_q^{\;k+l} = \sum_{r=0}^{k+l} \binom{k+l}{r} (-1)^{r} (1-p)^{rq}\]
so, substituting these into \ref{impJointPi} and combining like terms, we find an equation which is analogous to \ref{joint}
\[ Pr(Y=y|P=p)Pr(Z=z|P=p)Pr(P=p) = \]
\begin{equation}\label{impJoint}
\begin{split}
\frac{\binom{m}{y}\binom{n}{z}}{B(\alpha,\beta)} \sum_{i=0}^{y}\sum_{j=0}^{m-y} \sum_{k=0}^{z}\sum_{l=0}^{n-z} 
g(i,j,k,l) \sum_{r=0}^{k+l}  \binom{k+l}{r}(-1)^{r} p^{i+j + \alpha - 1 }(1-p)^{m - i - j + \beta - 1 + nq - kq - lq + rq}
\end{split}
\end{equation}
Now, by margining out $p$, we can find an equation which is analogous to \ref{margined}
\[ Pr(Y=y)Pr(Z=z) = \int_{0}^{1}Pr(Y=y|P=p)Pr(Z=z|P=p)Pr(P=p)dp \]
\[ = \frac{\binom{m}{y}\binom{n}{z}}{B(\alpha,\beta)} \sum_{i=0}^{y}\sum_{j=0}^{m-y} \sum_{k=0}^{z}\sum_{l=0}^{n-z} 
g(i,j,k,l) \sum_{r=0}^{k+l} \binom{k+l}{r}(-1)^{r} \int_{0}^{1} p^{i+j + \alpha - 1 }(1-p)^{m - i - j + \beta - 1 + nq - kq - lq + rq}dp \]
\begin{equation}\label{marginedImp}
\begin{split}
= \frac{\binom{m}{y}\binom{n}{z}}{B(\alpha,\beta)} \sum_{i=0}^{y}\sum_{j=0}^{m-y} \sum_{k=0}^{z}\sum_{l=0}^{n-z} 
g(i,j,k,l) \sum_{r=0}^{k+l} \binom{k+l}{r}(-1)^{r} B(i+j + \alpha, m - i - j + \beta + nq - kq - lq + rq)
\end{split}
\end{equation}
Using Bayes' Theorem \ref{bayes},
\[ Pr(P=p | Y=y, Z=z) \]
\begin{equation}\label{posteriorImp}
\begin{split}
= \frac
{\sum_{i=0}^{y}\sum_{j=0}^{m-y} \sum_{k=0}^{z}\sum_{l=0}^{n-z} g(i,j,k,l) \sum_{r=0}^{k+l}\binom{k+l}{r} (-1)^{r} p^{i+j + \alpha - 1 }(1-p)^{m - i - j + \beta - 1 + nq - kq - lq + rq}
}
{\sum_{i=0}^{y}\sum_{j=0}^{m-y} \sum_{k=0}^{z}\sum_{l=0}^{n-z} g(i,j,k,l) \sum_{r=0}^{k+l}\binom{k+l}{r} (-1)^{r} B(i+j + \alpha, m - i - j + \beta + nq - kq - lq + rq)}
\end{split}
\end{equation}
A similar approach can be used to create an analytical posterior distributions for $P$ under beta prior estimates of $s_e$ and $s_p$.

\section{Characterizing the Posterior Distribution for $P$}
In this section, we will consider only the case where $s_e = s_p = 1$.
\subsection{Moments of the Posterior Distribution for $P$}

The moments of the posterior distribution for $P$ can be calculated as follows. 
The $n$-th raw moment of a distribution, $\mu_n$, is defined as $\int x^{n} P(x)dx$, where $x$ is a random variable and P(x) is its PDF. Hence, for our application, the $n$-th moment of the posterior distribution for $P$ is:
\[ \mu_n = \int_{0}^{1}\frac {\sum_{i=0}^{z}  \binom{z}{i} (-1)^{i} B(\gamma, \delta + qi) f(p, \gamma, \delta + qi)} {\sum_{i=0}^{z}  \binom{z}{i} (-1)^{i} B(\gamma, \delta + qi)}p^{n}dp \]
Where, as before, $f(p, \gamma, \delta + qi)$ is the PDF of the beta distribution with parameters  $\gamma$ and $\delta + qi$, evaluated at $p$.
\[ f(p, \gamma, \delta + qi) = \frac{p^{\gamma-1}(1-p)^{\delta + qi-1}} { B(\gamma, \delta + qi) }\]
Using linearity of integration:
\[ \mu_n = \frac {\sum_{i=0}^{z}  \binom{z}{i} (-1)^{i} B(\gamma, \delta + qi) \int_{0}^{1} p^{n} f(p, \gamma, \delta + qi)dp} {\sum_{i=0}^{z}  \binom{z}{i} (-1)^{i} B(\gamma, \delta + qi)} \]
But, 
\[\int_{0}^{1} p^{n} f(p, \gamma, \delta + qi)dp\] is clearly just of the $n$-th moment of the Beta distribution $f(p, \gamma, \delta + qi)$, which is known to be 
\begin{equation}\label{moment_eqn_part}
\mu^i_n = \prod^n_{j=0} \frac{\gamma + j}{\gamma + \delta + qi + j}
\end{equation}
So the $n$-th moment of the posterior distribution for $P$ can be found using the equation
\begin{equation} \label{moment_eqn_full}
 \mu_n = \frac {\sum_{i=0}^{z}  \binom{z}{i} (-1)^{i} B(\gamma, \delta + qi) \mu^i_n} {\sum_{i=0}^{z}  \binom{z}{i} (-1)^{i} B(\gamma, \delta + qi)} 
\end{equation}
And thus the $n$-th raw moment of posterior distribution for $P$ is just a weighted sum of the $n$-th raw moments of its constituent beta distributions.

\subsection{The Posterior Distribution for $P$ is \underline{not} a Beta Distribution}
Since the posterior distribution for $P$ is a weighted sum of Beta distributions, it is natural to wonder if the posterior distribution for $P$ is itself another Beta distribution. We can construct a simple counterexample to show that the posterior distribution for $P$ is not in general a beta distribution. Take the case where $m = 1, y = 0, n = 1, z = 1, \alpha=1, \beta =1$ and $q = 3$. With these values, $\gamma = 1$ and $\delta = 2$. Plug these all into equation \ref{soln2}.
\[ Pr(P=p|Y=0, Z=1)  =  \frac {\sum_{i=0}^{1}  \binom{1}{i} (-1)^{i} B(1, 2 + 3i) f(p, 1, 2 + 3i)} {\sum_{i=0}^{1}  \binom{1}{i} (-1)^{i} B(1, 2 + 3i)}  \]
\[ = \frac {B(1,2)f(p,1,2)-B(1,5)f(p,1,5)} {B(1,2)-B(1,5)}\]
Note that $B(1,2)=\frac{1}{2}$ and $B(1,5)=\frac{1}{5}$, so
\begin{equation}\label{simple_case} Pr(P=p|Y=0, Z=1)  =  \frac {.5f(p,1,2)-.2f(p,1,5)} {.3} = \frac {5f(p,1,2)-2f(p,1,5)} {3} \end{equation}
Beta distributions are determined by their moments, so if this posterior distribution were a beta distribution, there would be some parameters $a$ and $b$ such that the moments of $f(p, a, b)$ match the moments of the posterior distribution for $P$. We will match the first three moments of $f(p, a, b)$ to those of the posterior distribution for $P$ to show that there is no solution for $a$ and $b$. First, we find the first three raw moments of the posterior distribution for $P$: $\mu_0$, $\mu_1$, and $\mu_2$. We will do this using equation \ref{moment_eqn_full}. 
From equation \ref{moment_eqn_part}, the moments of $f(p,1,2)$ are
\[ \mu^{0}_{0} = \frac{1}{1+2} = \frac{1}{3}\]
\[ \mu^{0}_{1} = (\frac{1}{1+2})(\frac{1+1}{1+2+1}) = \frac{1}{6}\]
\[ \mu^{0}_{2} = (\frac{1}{1+2})(\frac{1+1}{1+2+1})(\frac{1+2}{1+2+2}) = \frac{1}{10}\]
Similarly, the moments of $f(p,1,5)$ are
\[ \mu^{1}_{0} = \frac{1}{1+5} = \frac{1}{6}\]
\[ \mu^{1}_{1} = (\frac{1}{1+5})(\frac{1+1}{1+5+1}) = \frac{1}{21}\]
\[ \mu^{1}_{2} = (\frac{1}{1+5})(\frac{1+1}{1+5+1})(\frac{1+2}{1+5+2}) = \frac{1}{56}\]
Putting these six moments together with equations \ref{moment_eqn_full} and \ref{simple_case}, we get 
\[\mu_{0} 
= \frac {5\mu^{0}_{0}-2\mu^{1}_{0}} {3} 
= \frac {\frac{1}{3}5-\frac{1}{6}2} {3} 
=  \frac{4}{9}\]
\[\mu_{1} 
= \frac {5\mu^{0}_{1}-2\mu^{1}_{1}} {3} 
= \frac {\frac{1}{6}5-\frac{1}{21}2} {3} 
= \frac{31}{126} 
\]
\[\mu_{2} 
= \frac {5\mu^{0}_{2}-2\mu^{1}_{2}} {3} 
= \frac {\frac{1}{10}5-\frac{1}{56}2} {3} 
= \frac{13}{84}
\]
So the posterior distribution for $P$'s first three moments are $\frac{4}{9}$, $\frac{31}{126}$, and $\frac{13}{84}$. If the posterior distribution for $P$ were equivalent to a standard beta distribution $f(p, a, b)$, its first three moments could be calculated in terms of $a$ and $b$ using the equation for the $n$-th moment of the beta distribution. Set these equations for the first three moments of  $f(p, a, b)$ equal to the moments we just calculated. 
\[ \mu_{0}^{*} = \frac{a}{a+b} = \frac{4}{9}\]
\[ \mu_{1}^{*} = (\frac{a}{a+b}) (\frac{a+1}{a+b+1}) = \frac{31}{126}\]
\[ \mu_{2}^{*} = (\frac{a}{a+b}) (\frac{a+1}{a+b+1}) (\frac{a+2}{a+b+2}) = \frac{13}{84} \]
There is no solution $(a,b)$ which satisfies these equations, so the posterior distribution for $P$ is not in general a beta distribution.

\subsection{Implementation Notes}
The coefficients of the summations can be calculated using a computer program. Some of the terms in both the numerator and denominator sums are very large or very small. Therefore, computer implementations of this distribution must use floating point algebra of sufficient precision. The python decimal package with 200 places of precision is usually precise enough to accurately compute confidence intervals. The CDF can also be calculated by computer. It is simply a weighted sum of the CDFs of the $1+z$ beta distributions. The same is true for the MGF of this distribution. This solution may be approximated by another beta distribution in some cases, but it is not a beta distribution in general. The solution can be approximated by assuming that it is equivalent to a beta distribution and using the method of moments, but this is only a good approximation if the true $P$ is small. 

\subsection{Sample Posterior Probability Distributions}
\begin{figure}[H]
\centering
\begin{subfigure}{.4\textwidth}
  \centering
  \includegraphics[width=1\linewidth]{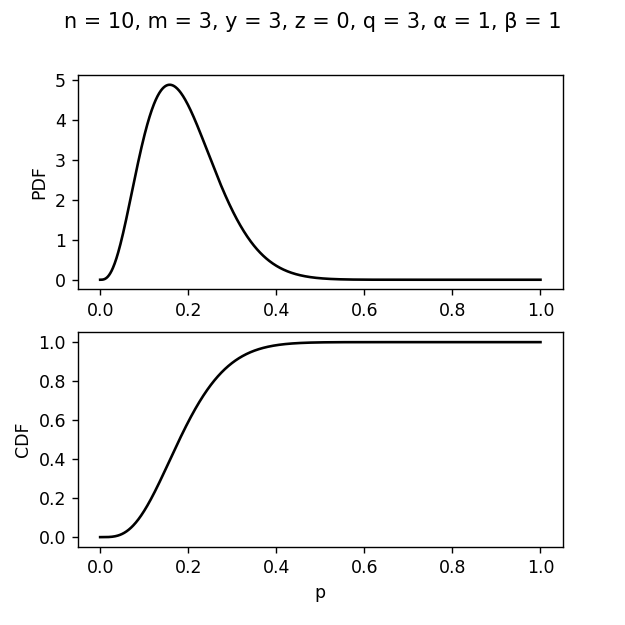}
  \caption{}
  \label{fig:sub1}
\end{subfigure}%
\begin{subfigure}{.4\textwidth}
  \centering
  \includegraphics[width=1\linewidth]{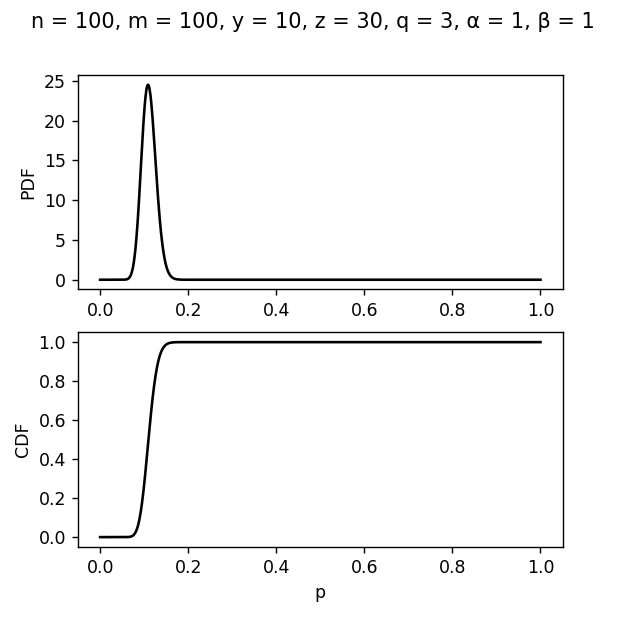}
  \caption{}
  \label{fig:sub2}
\end{subfigure}
\begin{subfigure}{.4\textwidth}
  \centering
  \includegraphics[width=1\linewidth]{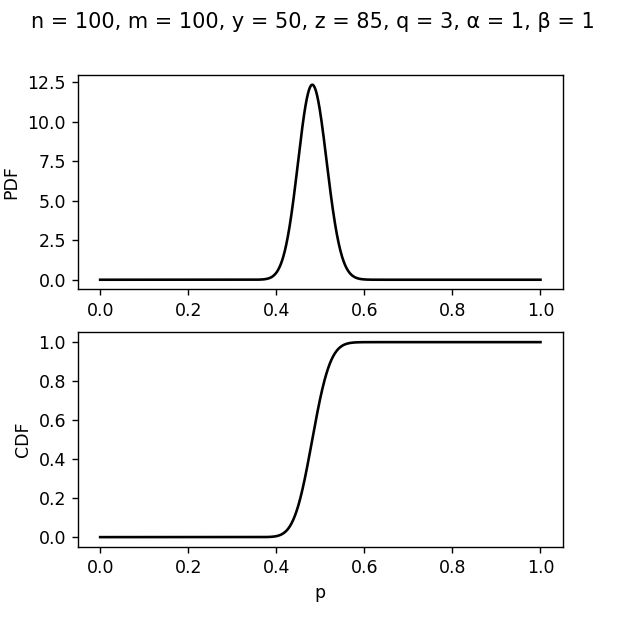}
  \caption{}
  \label{fig:sub3}
\end{subfigure}
\begin{subfigure}{.4\textwidth}
  \centering
  \includegraphics[width=1\linewidth]{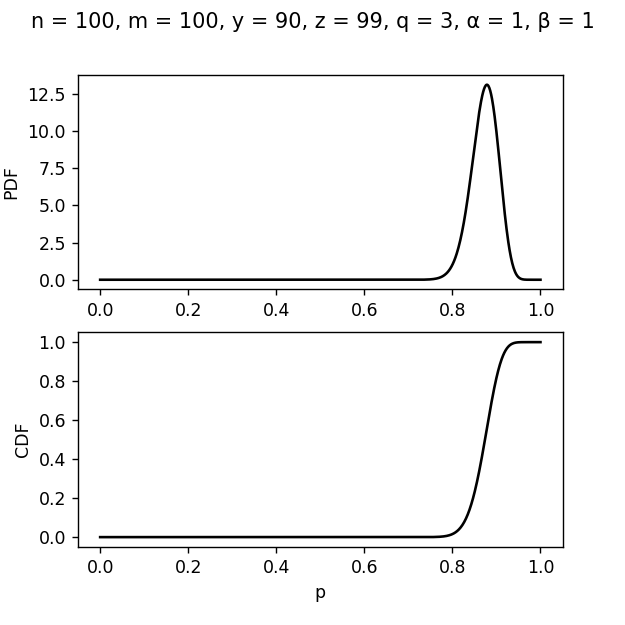}
  \caption{}
  \label{}
\end{subfigure}
\caption{Example Posterior Distributions for $P$}
\label{fig:ex}
\end{figure}

Figure \ref{fig:ex} shows example posterior probability distributions for $P$, computed following the equation above. The parameter settings used to create a curve are displayed above the graphs.

\section{Simulation 1: $s_e = s_p = 1$}
\subsection{Simulation Design}

For Simulation 1, we assume that $s_e=s_p=1$. We wrote a computer program to simulate $m$ individual tests and $n$ pooled tests with $q$ individuals per pooled test and variable true population prevalence $P$. We can use the results of these simulated tests to construct the analytical posterior probability distribution for $P$, calculate $95\%$ confidence intervals for $P$, and compute various properties of the posterior distribution for $P$. 

To assess the method's performance across a range of true prevalence values, we varied the true $P$ between 0.01 and 0.99 in increments of approximately 0.05, or each of the following: \\
$P \in [0.01, 0.05, 0.1 , 0.15, 0.2 , 0.25, 0.3 , 0.35, 0.4 , 0.45, 0.5 , 0.55, 0.6, 0.65, 0.7 , 0.75, 0.8 , 0.85, 0.9 , 0.95, 0.99]$.\\
To assess the method's performance across a range of numbers of individual and pooled tests, we varied $m$, the number of individual tests, between 0 and 200 in increments of 20, or each of the following:\\
$m \in [0,20,40,60,80,100,120,140,160,180,200]$.\\
For each trial, the number of pooled tests $n$ was $200 - m$ so that we always simulated 200 tests total.\\
We varied $q$, the number of individuals per pooled test, between 3 and 6 inclusive in increments of 1.\\
We ran 100 trials per experimental condition, or combination of $(m, n, P, q)$, for a total of 92400 trials. For each trial, we recorded: 
\begin{enumerate}
\setlength{\itemindent}{5em}
\item if the true $P$ was inside our 95\% confidence interval, 
\item the width of the confidence interval, and
\item the expected value of the posterior distribution for $P$.
\end{enumerate}
Then, the results for each of these measurements were aggregated for each experimental condition  $(m, n, P, q)$. 

\subsection {Simulation Results}

\subsubsection {Confidence Interval Accuracy}
If our derivation is correct, the posterior distribution for $P$'s 95\% confidence interval for $P$ will contain the true value of $P$ about 95\% of the time. The data shows that this holds true.
We now calculate the confidence interval accuracy for all experiments. We ran 100 simulated trials for each parameter setting. Each point in this graph shows the proportion of trials at a single experimental condition in which the posterior distribution for $P$'s 95\% confidence interval for $P$ will contained the true value of $P$. We separate the figures by value of $q$ for legibility.
\begin{figure}[H]\centering\includegraphics[scale=0.45]{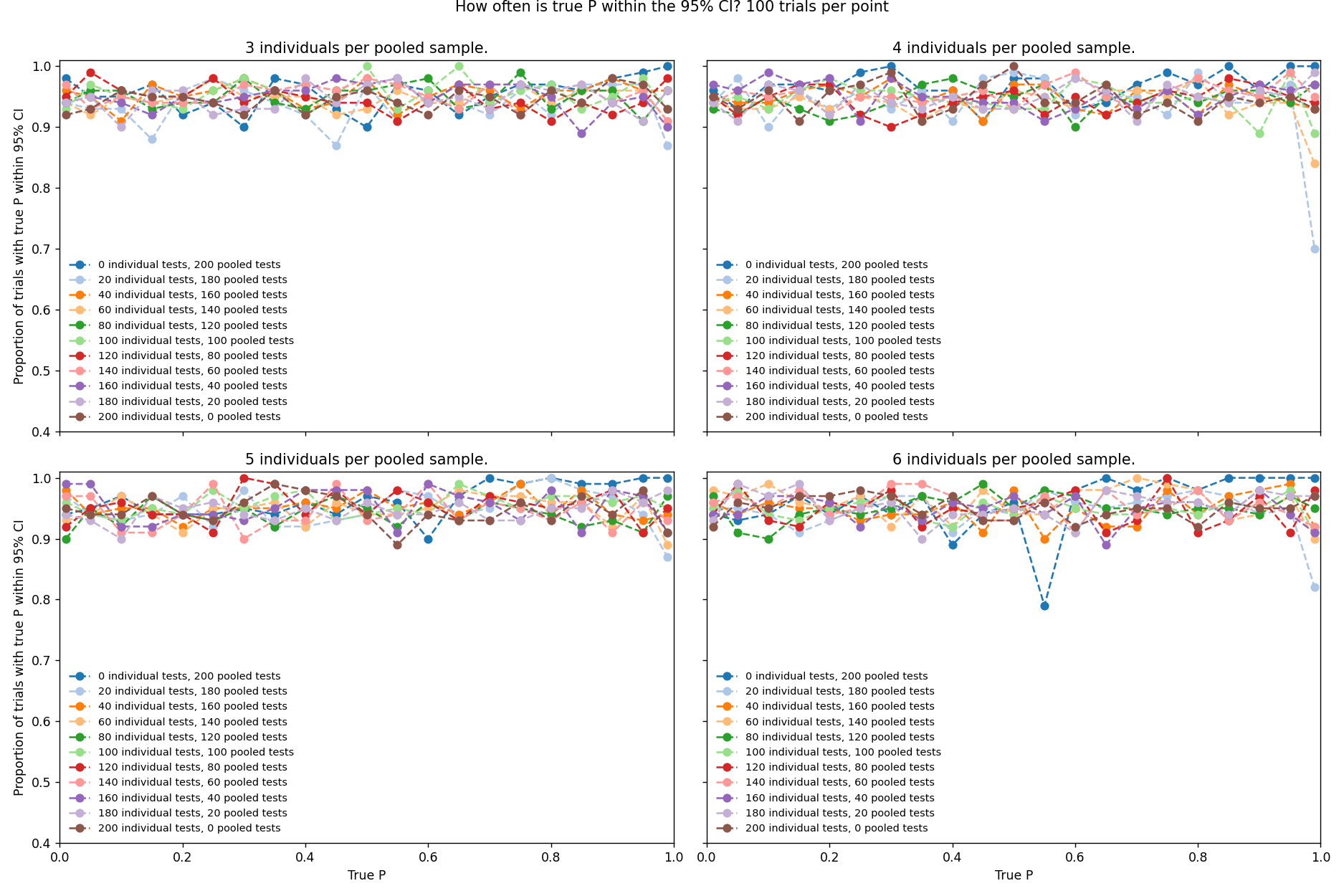}\caption{Confidence Interval Accuracies}\end{figure}
\subsubsection {Confidence Interval Width}
We define $f(p)$ as the posterior distribution for $P$ and $F(p)$ as the CDF of this distribution.
We define the 95\% Confidence Interval Width as 
\[ {CI}_{width} = F^{-1}(0.975) - F^{-1}(0.025)\]
${CI}_{width}$ decreases if we have increased confidence in our estimate of the true value of $P$.
We calculate the 95\% confidence interval widths for all experiments. We ran 100 simulated trials for each parameter setting. Each point in this graph shows the mean and standard deviation of the CI width, aggregated over 100 trials run for each parameter setting. We separate the figures by value of $q$ for legibility.
\begin{figure}[H]\centering\includegraphics[scale=.45]{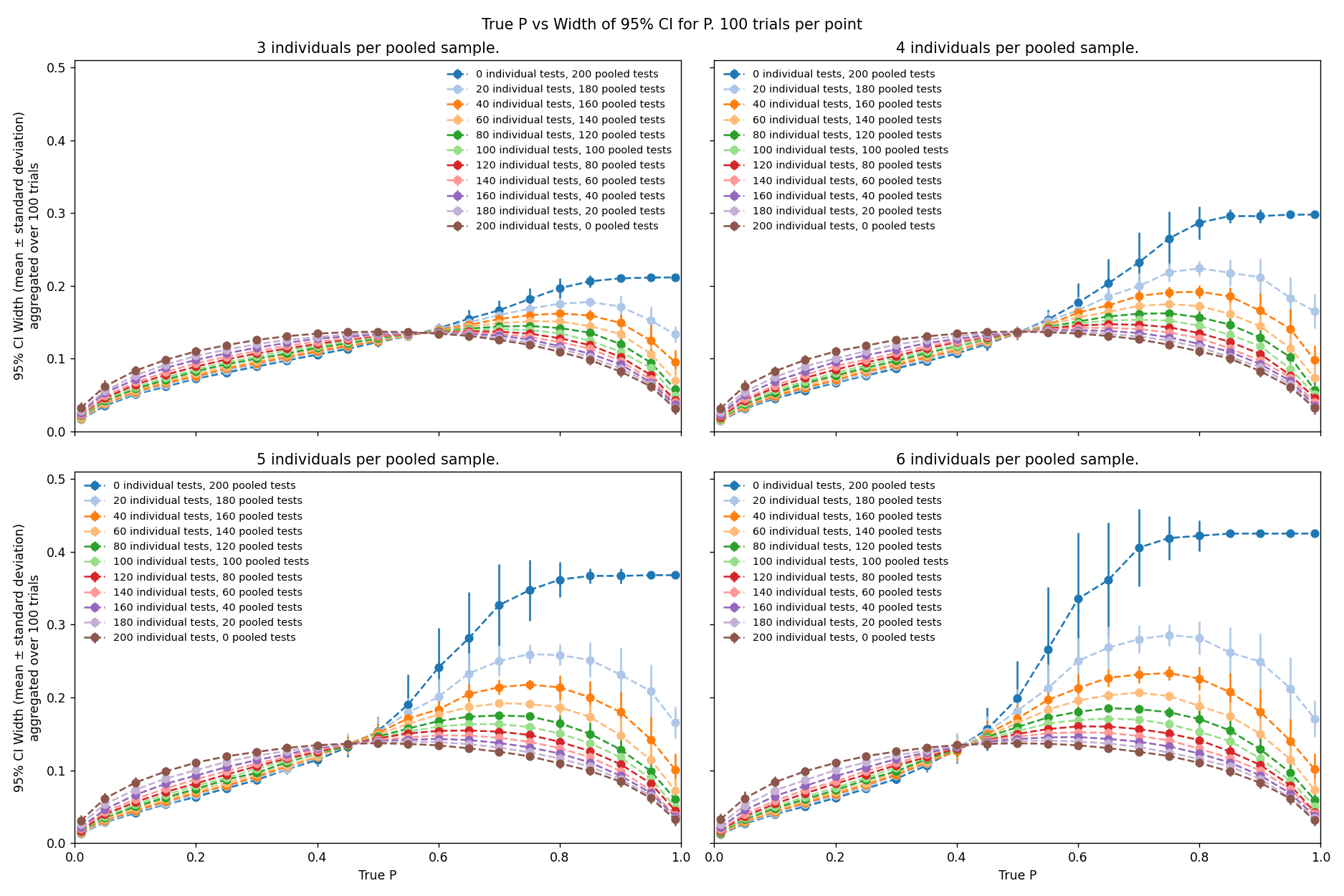}\caption{Confidence Interval Widths}\end{figure}
\begin{figure}[H]\centering\includegraphics[scale=0.5]{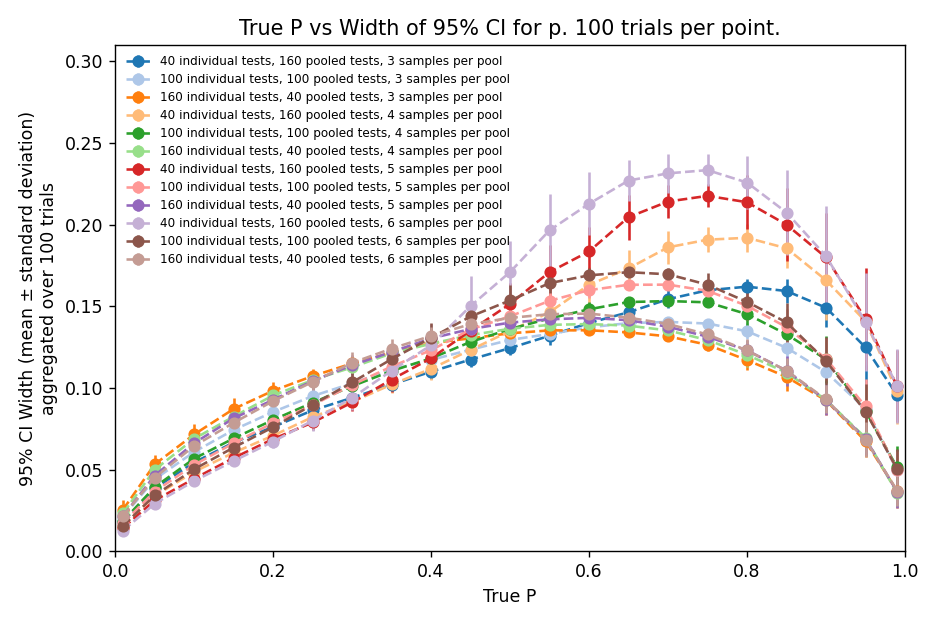}\caption{Confidence Interval Widths for multiple values of $q$}\end{figure}
\begin{figure}[H]\centering\includegraphics[scale=0.45]{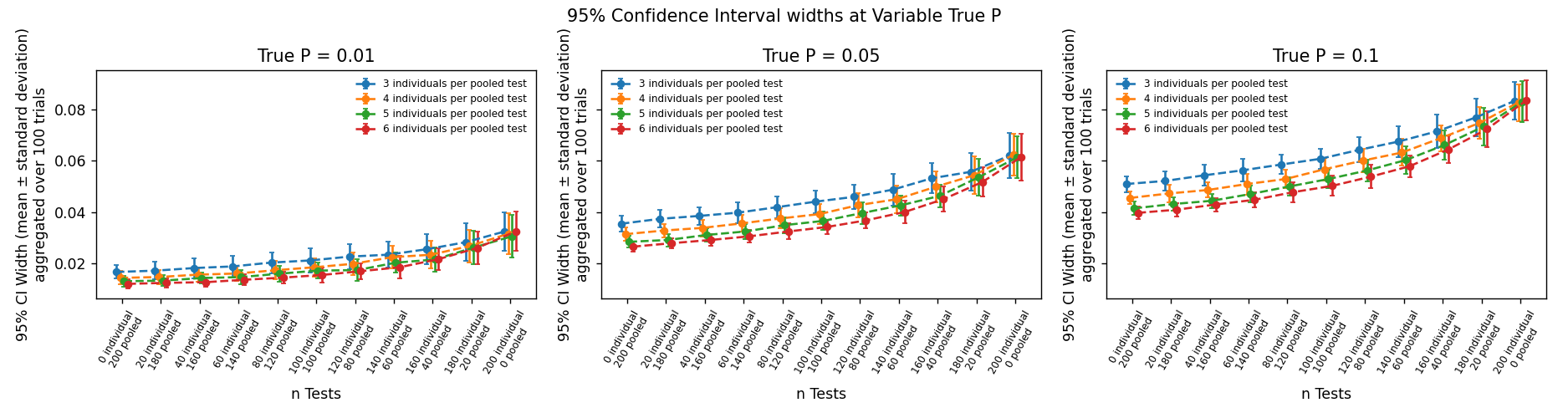}\caption{Confidence Interval Widths By Sampling Design}\end{figure}
\begin{figure}[H]\centering\includegraphics[scale=0.6]{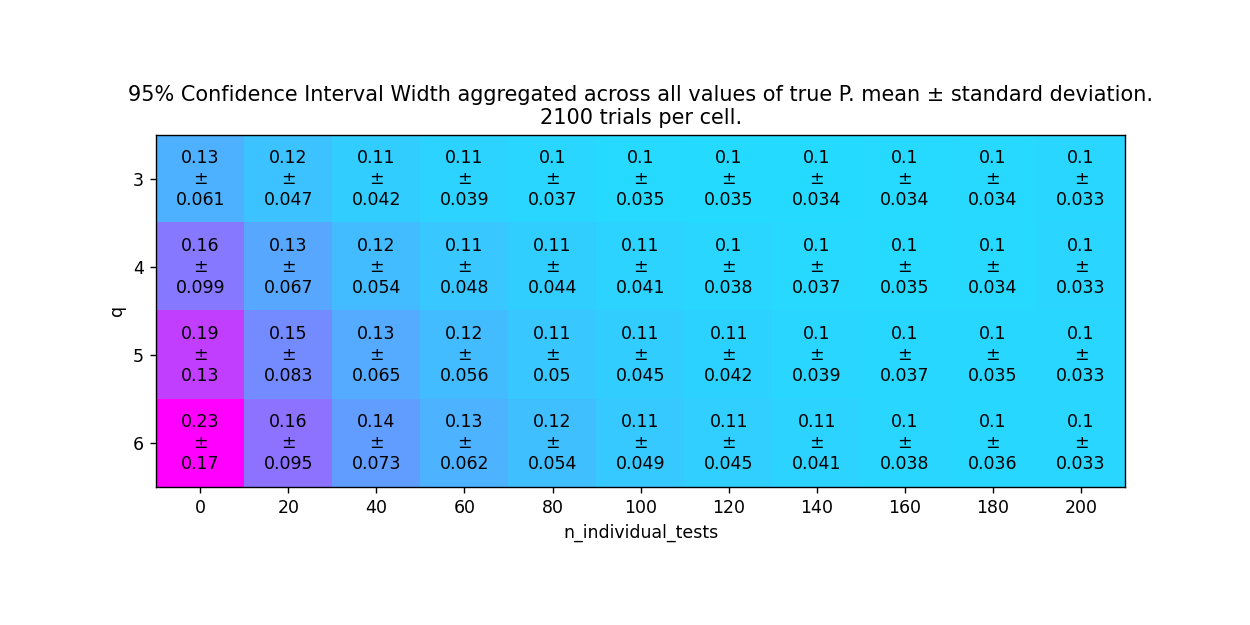}\caption{Confidence Interval Width aggregated across all true $P$}\end{figure}
\subsubsection {Expected Value of the Posterior Distribution for $P$}
By definition of Expectation,
\[ E[P=p|Y=y, Z=z] = \int_{0}^{1}Pr(P=p|Y=y, Z=z)dp\]\\
We now calculate $E[P=p|Y=y, Z=z]$ for all experimental trials. We ran 100 simulated trials for each parameter setting. Each point in this graph shows the mean and standard deviation of $E[P=p|Y=y, Z=z]$, aggregated over 100 trials run for each parameter setting
\begin{figure}[H]\centering\includegraphics[scale=0.45]{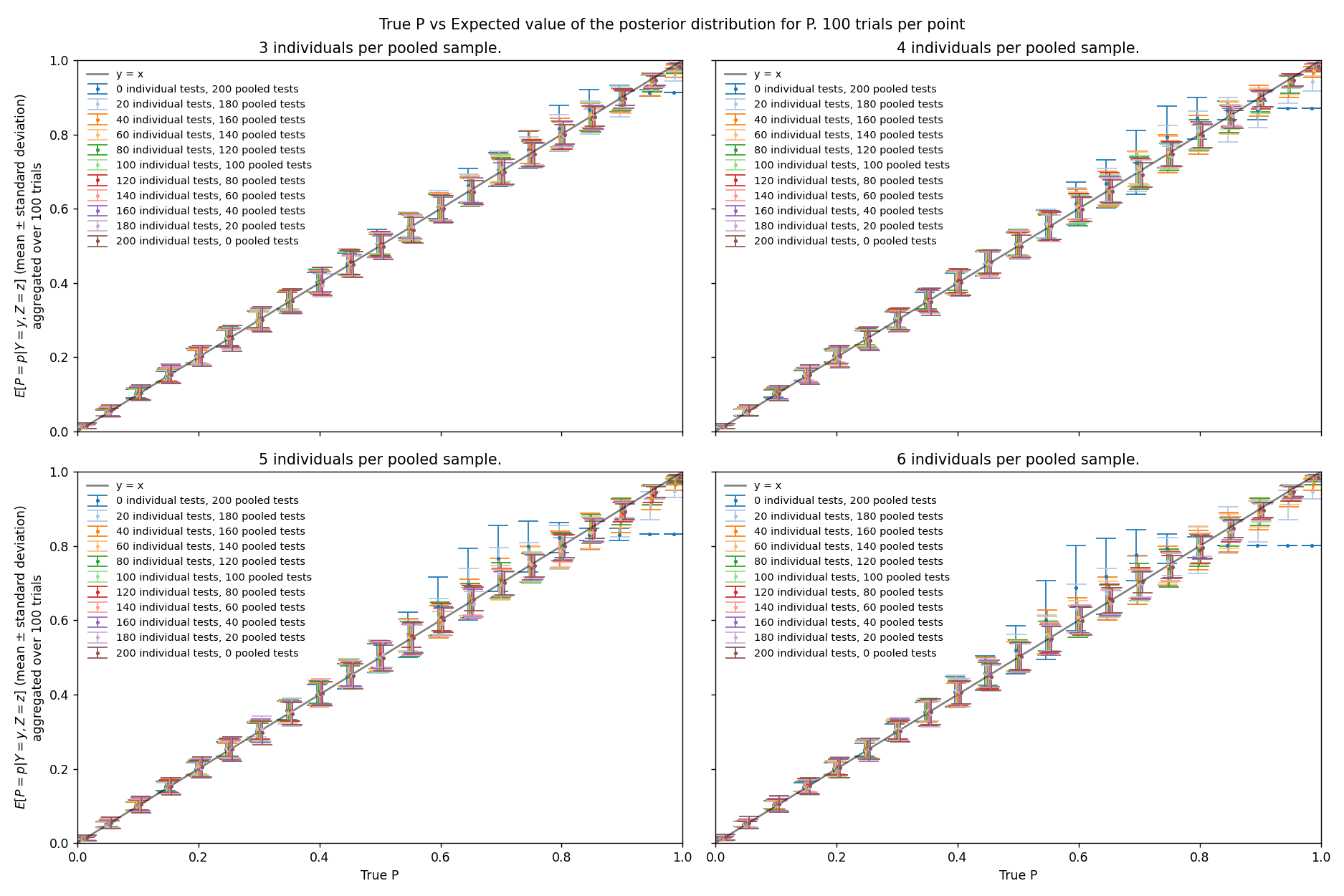}\caption{$E[P=p|Y=y, Z=z]$}\end{figure}
We calculate the percent error of expectation as follows:
\[ \% Error = \frac{|E[P=p|Y=y, Z=z] - P_{true}|}{P_{true}}\]
Where $P_{true}$ is true population prevalence.
\begin{figure}[H]\centering\includegraphics[scale=0.6]{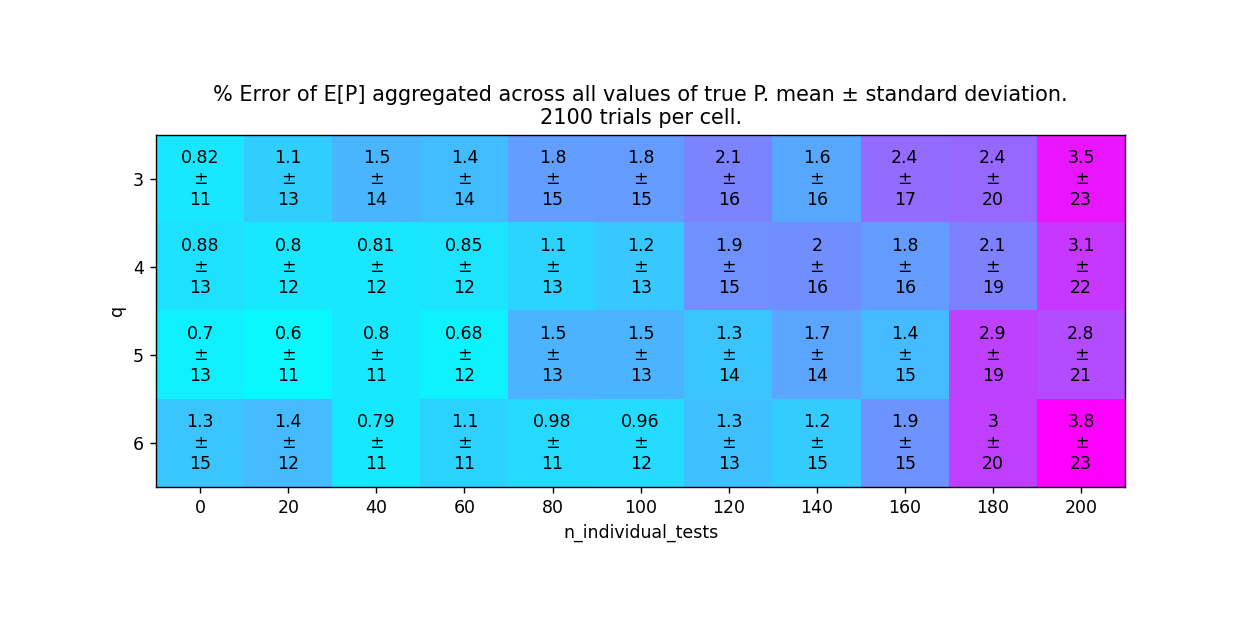}\caption{Percent error of $E[P=p|Y=y, Z=z]$ aggregated for each $q$ and number of individual tests ($m$)}\end{figure}

\section{Simulation 2: $s_e = s_p < 1$}
\subsection{Simulation Design}
For Simulation 2, we used the results of \ref{sesp} to compute analytical posterior distributions for $P$ under conditions where testing sensitivity and specificity are not equal to 1. We wrote a computer program to simulate $m$ individual tests and $n$ pooled tests with $q$ individuals per pooled test and variable true population prevalence $P$. We can use the results of these simulated tests to construct the analytical posterior probability distribution for $P$, calculate $95\%$ confidence intervals for $P$, and compute various properties of the posterior distribution for $P$. We simulated a smaller number of tests for each condition and a reduced variety of conditions because of the increased computational costs of inference under imperfect testing.

To assess the method's performance across a range of true prevalence values, we varied the true $P$ between 0.01 and 0.99 in increments of approximately 0.05, or each of the following:\\
$P \in [0.01, 0.05, 0.1 , 0.15, 0.2 , 0.25, 0.3 , 0.35, 0.4 , 0.45, 0.5 , 0.55, 0.6, 0.65, 0.7 , 0.75, 0.8 , 0.85, 0.9 , 0.95, 0.99]$.\\
To assess the method's performance across a range of numbers of individual and pooled tests, we varied $m$, the number of individual tests, between 0 and 30 in increments of 5, or each of the following:\\
$m \in [0, 5, 10, 15, 20, 25, 30]$.\\
For each trial, the number of pooled tests $n$ was $30 - m$ so that we always simulated 30 tests total. \\
We assumed that we know the true values of $s_e$ and $s_p$. We varied $s_e$ and $s_p$ between each of the following:\\
$s_e = s_p = 1$, $s_e = s_p = 0.95$, $s_e = s_p = 0.9$, and $s_e = s_p = 0.8$.\\
We used $q=3$ for all testing conditions.\\
We ran 100 trials per experimental condition, or combination of $(m, n, P, q, s_e, s_p)$, for a total of 58800 trials.\\
For each trial, we recorded: 
\begin{enumerate}
\setlength{\itemindent}{5em}
\item if the true $P$ was inside our 95\% confidence interval, 
\item the width of the confidence interval, and
\item the expected value of the posterior distribution for $P$.
\end{enumerate}
Then, the results for each of these measurements were aggregated for each experimental condition  $(m, n, P, q, s_e, s_p)$. 

\subsection {Simulation Results}

\subsubsection {Confidence Interval Accuracy}

If our derivation is correct, the posterior distribution for $P$'s 95\% confidence interval for $P$ will contain the true value of $P$ about 95\% of the time. The data shows that this holds true.

We now calculate the confidence interval accuracy for all experiments. We ran 100 simulated trials for each parameter setting. Each point in this graph shows the proportion of trials at a single experimental condition in which the posterior distribution for $P$'s 95\% confidence interval for $P$ will contained the true value of $P$. We separate the figures by value of $s_e$ and $s_p$ for legibility.
\begin{figure}[H]\centering\includegraphics[scale=0.45]{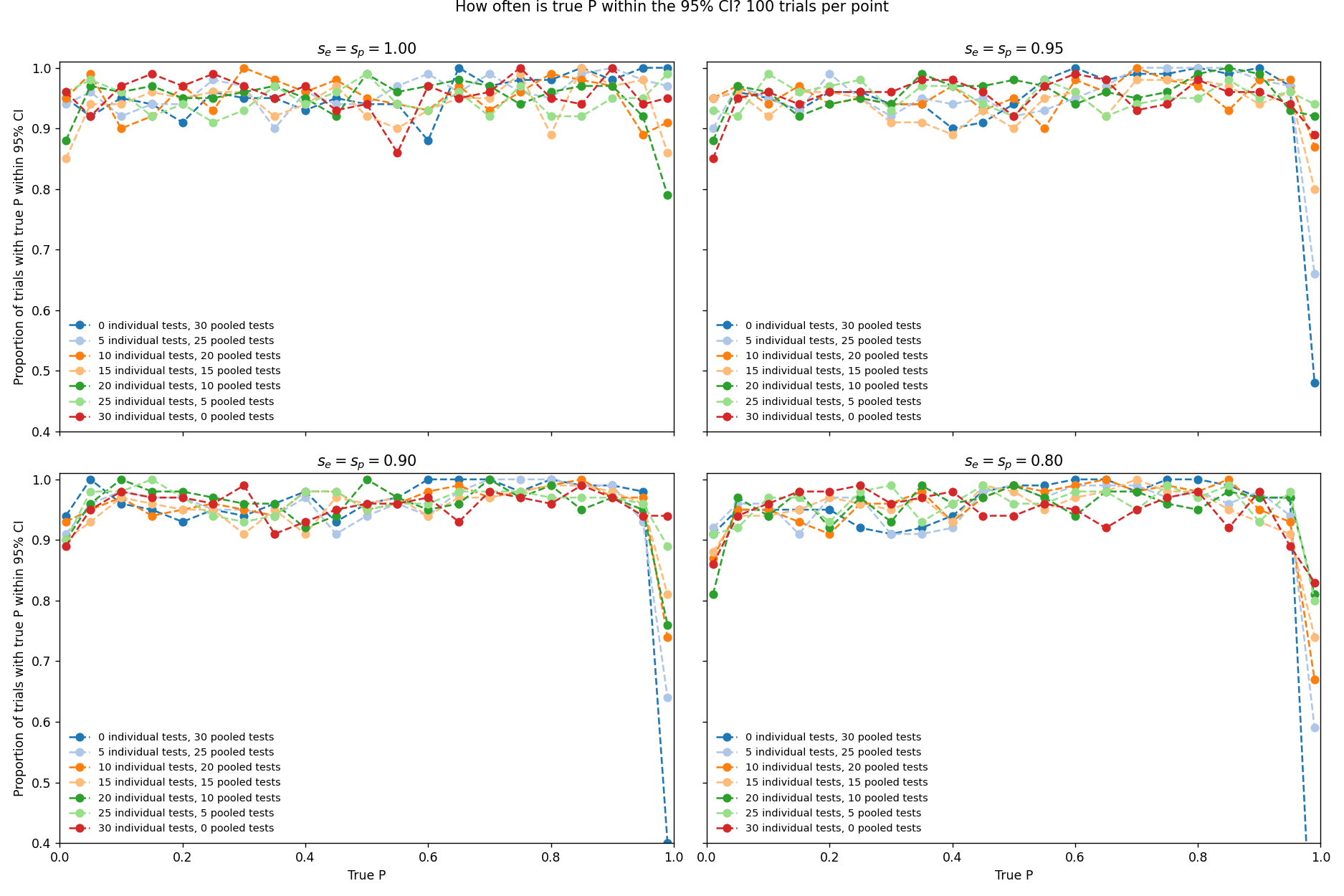}\caption{Confidence Interval Accuracies}\end{figure}

\subsubsection {Confidence Interval Width}
We define $f(p)$ as the posterior distribution for $P$ and $F(p)$ as the CDF of this distribution.
We define the 95\% Confidence Interval Width as 
\[{CI}_{width} = F^{-1}(0.975) - F^{-1}(0.025)\]
${CI}_{width}$ decreases if we have increased confidence in our estimate of the true value of $P$. We calculate the 95\% confidence interval widths for all experiments. We ran 100 simulated trials for each parameter setting. Each point in this graph shows the mean and standard deviation of the CI width, aggregated over 100 trials run for each parameter setting. We separate the figures by value of $s_e$ and $s_p$ for legibility.
\begin{figure}[H]\centering\includegraphics[scale=0.45]{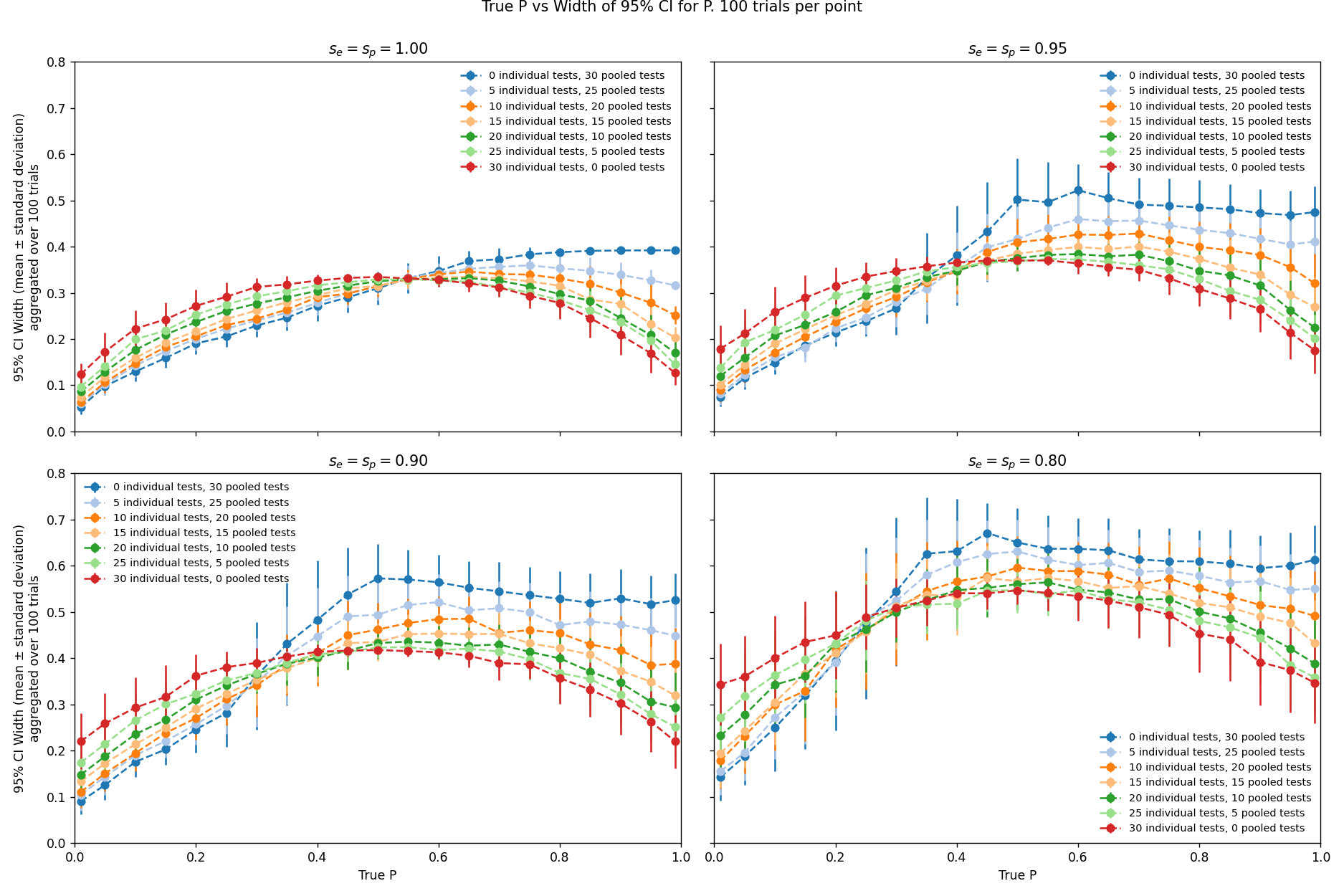}\caption{Confidence Interval Widths}\end{figure}
\begin{figure}[H]\centering\includegraphics[scale=0.5]{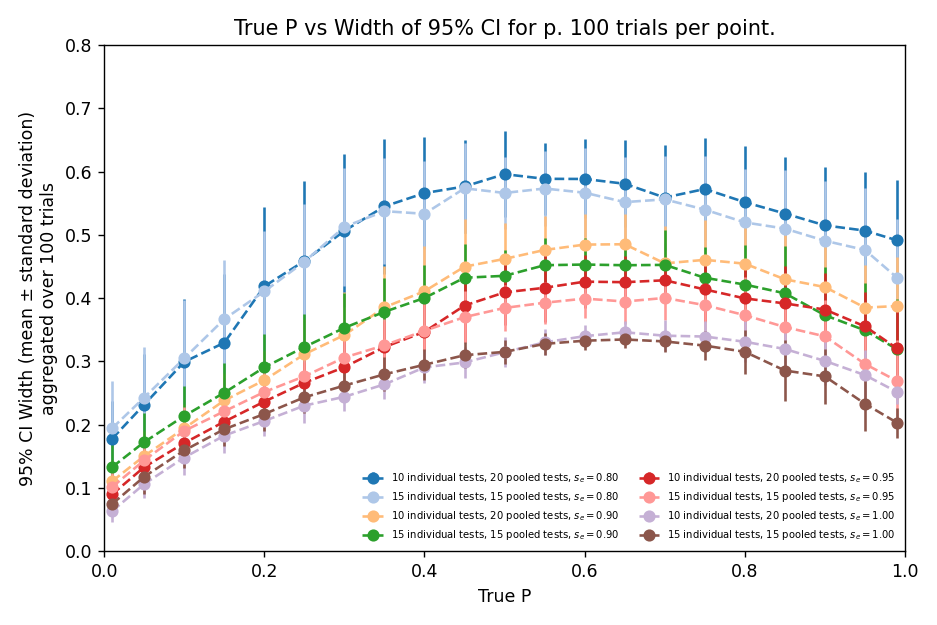}\caption{Confidence Interval Widths for multiple values of $s_e$ and $s_p$}\end{figure}
\begin{figure}[H]\centering\includegraphics[scale=0.45]{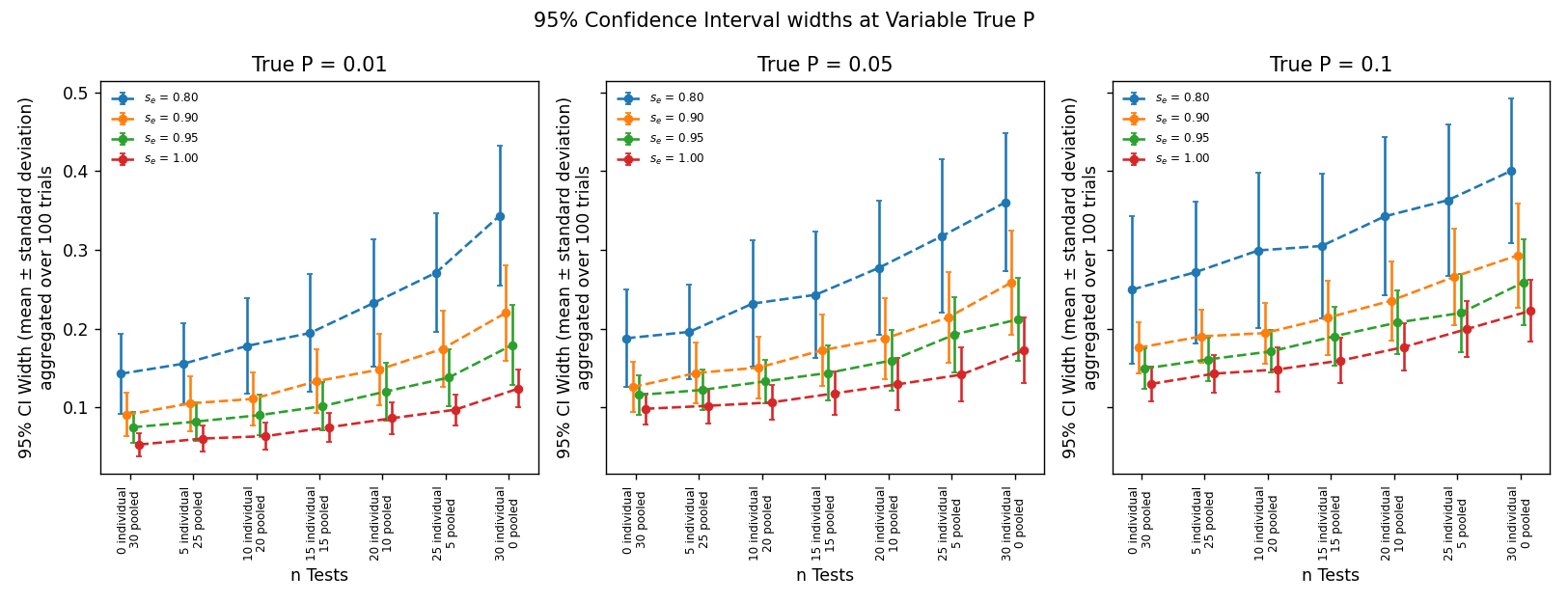}\caption{Confidence Interval Widths By Sampling Design}\end{figure}
\begin{figure}[H]\centering\includegraphics[scale=0.6]{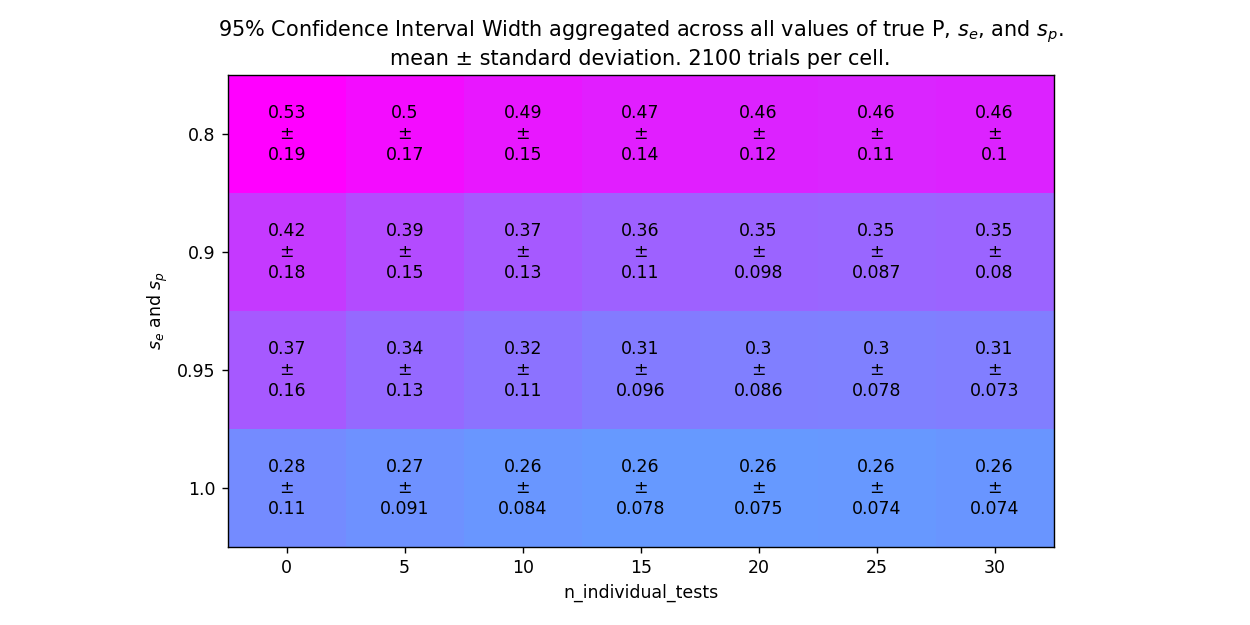}\caption{Confidence Interval Width aggregated across all true $P$ and values of $s_e$ and $s_p$}\end{figure}

\subsubsection {Expected Value of the Posterior Distribution for $P$}
Again, by definition of Expectation,
\[ E[P=p|Y=y, Z=z] = \int_{0}^{1}Pr(P=p|Y=y, Z=z)dp\]
We now calculate $E[P=p|Y=y, Z=z]$ for all experimental trials. We ran 100 simulated trials for each parameter setting. Each point in this graph shows the mean and standard deviation of the expected values aggregated over 100 trials run for each parameter setting. We separate the figures by value of $s_e$ and $s_p$ for legibility.
\begin{figure}[H]\centering\includegraphics[scale=0.45]{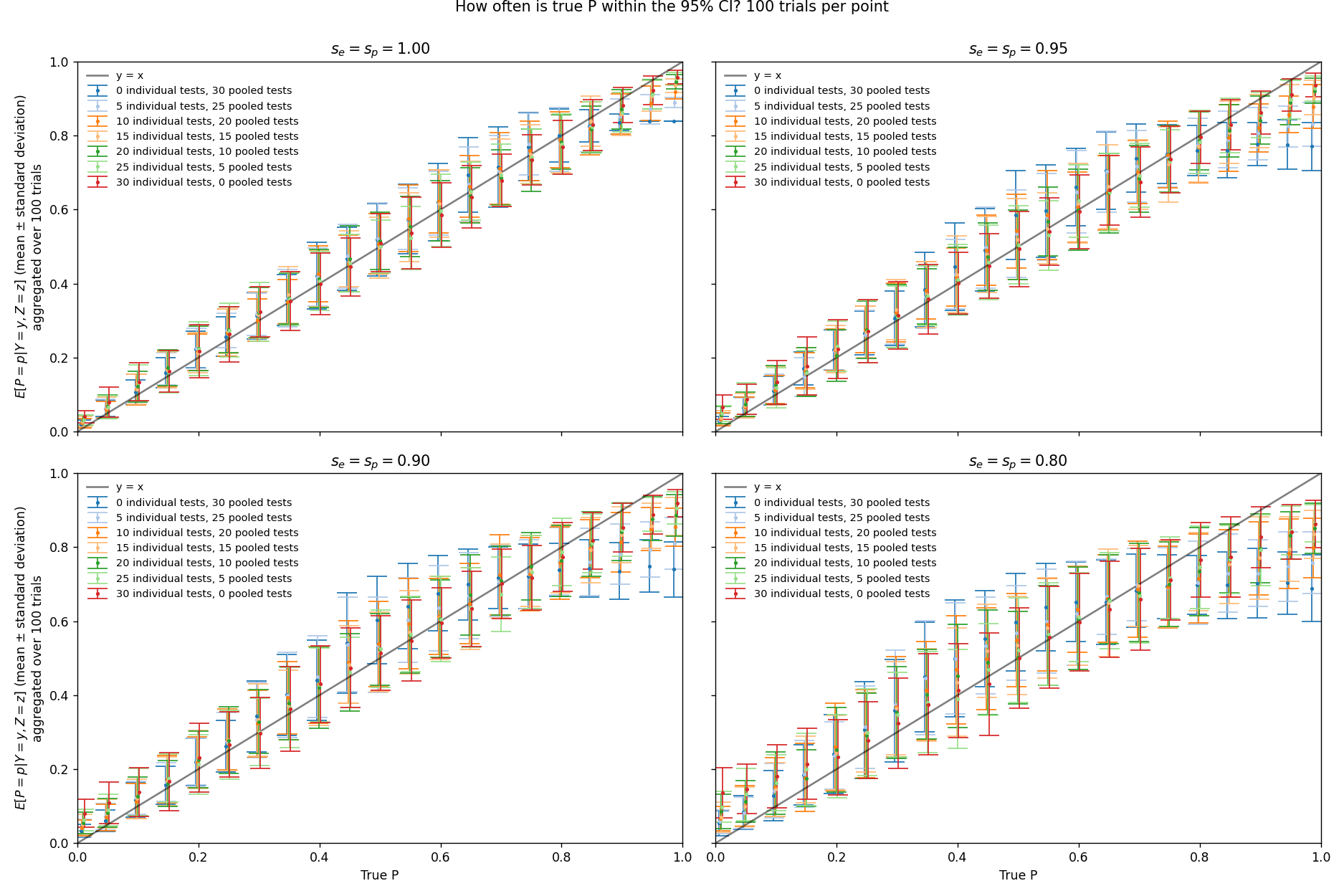}\caption{$E[P=p|Y=y, Z=z]$}\end{figure}
We calculate the percent error of expectation as $\% Error = \frac{|E[P=p|Y=y, Z=z] - P_{true}|}{P_{true}}$, where $P_{true}$ is true population prevalence.
\begin{figure}[H]\centering\includegraphics[scale=0.6]{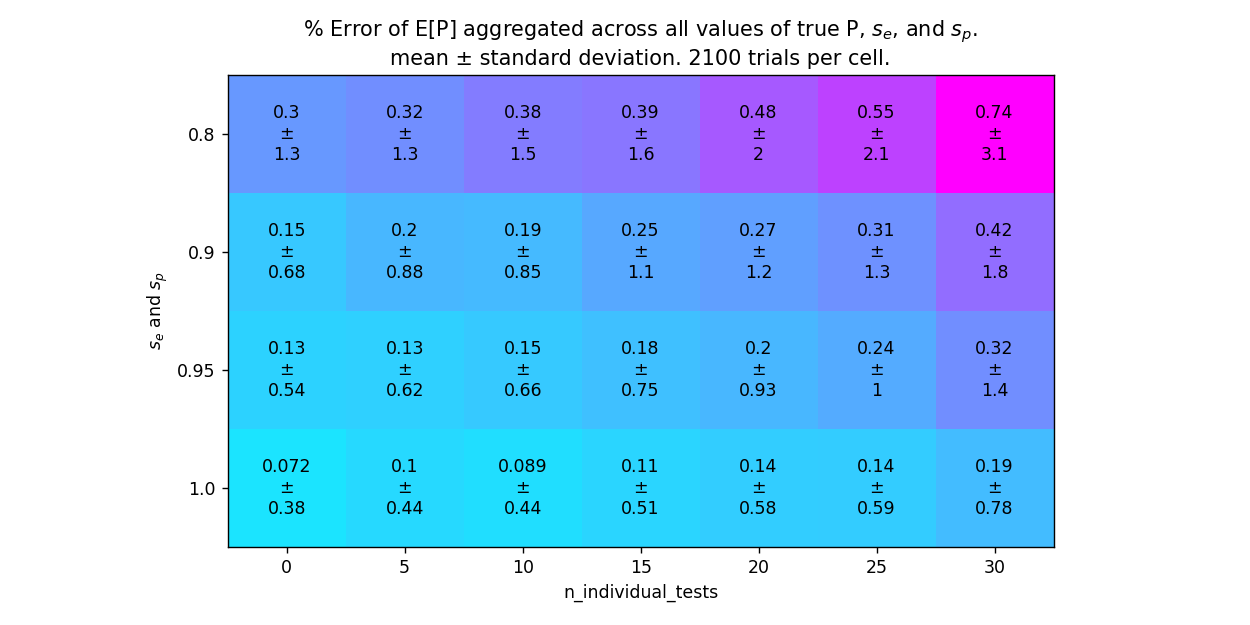}\caption{Percent error of $E[P=p|Y=y, Z=z]$ aggregated for each value of $s_e$ and $s_p$ and number of individual tests ($m$)}\end{figure}

\section{Discussion}

We have presented an analytical method for estimating population prevalence from combined individual and pooled binary sampling data. We have also conducted simulations to characterize these posterior distributions under a variety of sampling conditions, including a range of true prevalences, variable numbers of pooled and individual tests, variable number of individual samples per pooled sample, and a range of values for test sensitivity and specificity. 

We computed the proportion of trials with the true $P$ falling within the confidence interval to see if in general the posterior probability distribution for $P$ accurately captures the true $P$. With $n_{trials}$ total trials conducted, the number of trials falling within a 95\% confidence interval should be distributed as $Binomial(n_{trials}, 0.95)$. Under almost all parameter settings, the proportion of trials with the true $P$ falling within the 95\% confidence interval for $P$ are consistent with expectations. The exceptions are some of the trials conducted with many pooled tests and a true $P$. Pooled tests conducted with a high true p are almost guaranteed to be positive, so they yield very little information. The increased inaccuracy in this high true $P$, high numbers of pooled tests region is due to this “washing out” effect. In general, performance degrades when true $P$ is greater than 0.95 or less than 0.05. Performance also degrades as sensitivity and specificity decrease, but our posterior distribution for $P$ still captures the truth under many parameter settings.

We computed the expected value of the posterior distributions for $P$ to assess if the posterior distribution can furnish us an adequate point estimate of $P$. The comparisons of true $P$ vs the expected value of the posterior distribution for $P$ (predicted $P$) follow similar trends as we saw with the confidence intervals. Predicted $P$ accuracy and precision are very good in almost all cases other than trials run with high true $P$, high numbers of pooled tests, and high $q$. Our results suggest that with larger $q$, $s_e$, and $s_p$, the expected value will overestimate true $P$ when its true value is close to 0 and underestimate $P$ when its true value is close to 1.

We computed confidence interval width to assess the precision of the posterior distribution for $P$. A wider confidence interval can be interpreted as greater uncertainty in the estimate of $P$. Using more pooled tests and more individuals per pooled test yields narrower confidence interval widths at low population prevalences but wider confidence interval widths at high population prevalences. In addition, the individual and pooled sample data both follow binomial distributions and are therefore at their highest variance around $P = 0.5$ or $\pi_q = 0.5$. Thus, as true $P$ increases, CI width increases to a maximum and then decreases.

Overall, these results show that this method is performant in all but extreme sampling conditions.

\printbibliography[heading=bibnumbered]

\end{document}